\documentclass[a4paper,12pt,eqno]{article}
\usepackage{theorem}
\usepackage{latexsym,amssymb,amsfonts,amsmath}
\usepackage{graphicx}
\newtheorem{thm}{Theorem}[section]
\newtheorem{lem}[thm]{Lemma}
\newtheorem{cor}[thm]{Corollary}

\theoremheaderfont{\scshape}

\newcommand{\ZM}{\mathbb{Z}}

\newcommand{\CM}{\mathbb{C}}

\newcommand{\ket}[1]{|#1\rangle}

\title{{\Large {\bf Sojourn times of the Hadamard walk \\ in one dimension}}}
\author{
{\small Norio Konno}\\
{\scriptsize Department of Applied Mathematics, 
Faculty of Engineering, 
Yokohama National University}\\
{\scriptsize Hodogaya, Yokohama 240-8501, Japan}\\
{\scriptsize e-mail: konno@ynu.ac.jp}\\
}
\date{\empty }
\begin{document}
\maketitle

\par\noindent
\begin{small}
\par\noindent
{\bf Abstract}. The Hadamard walk is a typical model of the discrete-time quantum walk. We investigate sojourn times of the Hadamard walk on a line by a path counting method.

\footnote[0]{
{\it Abbr. title:} Sojourn times of the Hadamard walk
}
\end{small}

\setcounter{equation}{0}
\section{Introduction}
Classical random walks are very useful tools for various fields. Their interesting properties have been extensively studied. In particular, some results on sojourn times of the discrete-time random walk in one dimension are counter-intuitive. For instance, the following two examples are well known, see \cite{Feller1968}. Let $c^{(2n)} (2k)$ denote the probability that in the time interval from 0 to $2n$ the walker starting from the origin spends $2k \> (0 \le k \le n)$ time units on the positive side. Then the probability distribution $\{c^{(2n)} (2k) : k=0,1, \ldots ,n \}$ is governed by the discrete arc sine law (see Section 3 below). One feels intuitively that $c^{(2n)} (2k)$ might have the greatest value at the central term, that is, $k=n/2$ for $n=0 \>$ (mod 2), or $k=(n-1)/2$ and $=(n+1)/2$ for $n=1 \>$ (mod 2). However the opposite is true, i.e., $c^{(2n)} (2k)$ has the greatest value at $k=0$ and $k=n$. Moreover $\tilde{c}^{(2n)} (2k)$ denotes the probability that in the time interval from 0 to $2n$ the walker starting from the origin and returning to this point spends $2k \> (0 \le k \le n)$ time units on the positive side. Contrary to intuition, the probability distribution $\{\tilde{c}^{(2n)} (2k) : k=0,1, \ldots ,n \}$ is independent of $k$, that is, $\tilde{c}^{(2n)} (2k) = 1/(n+1) \> (k=0,1, \ldots ,n)$, see Section 3 below. The quantum walk (QW) is a quantum analog of the classical random walk. The corresponding results on sojourn times of the one-dimensional discrete-time QW are not known. So we give the quantum counterparts by a path counting approach. Reviews and a book on QWs are Kempe \cite{Kempe2003}, Kendon \cite{Kendon2007}, Venegas-Andraca \cite{VAndraca2008}, Konno \cite{Konno2008b}. 

The rest of the paper is organized as follows. Section 2 gives the definition of our model. In Sect. 3, we present our main results (Theorems {\rmfamily \ref{thm1a}} and {\rmfamily \ref{thm2a}}) of this paper. Section 4 is devoted to the proof of Theorem {\rmfamily \ref{thm1a}}. In Sect. 5, we prove Theorem {\rmfamily \ref{thm2a}}.

\section{Definition of the walk}
In this section, we give the definition of the two-state QW on $\ZM$ considered here, where $\ZM$ is the set of integers. The discrete-time QW is a quantum version of the classical random walk with an additional degree of freedom called chirality. The chirality takes values left and right, and it means the direction of the motion of the walker. At each time step, if the walker has the left chirality, it moves one step to the left, and if it has the right chirality, it moves one step to the right. In this paper, we put
\begin{eqnarray*}
\ket{L} = 
\left[
\begin{array}{cc}
1 \\
0  
\end{array}
\right],
\qquad
\ket{R} = 
\left[
\begin{array}{cc}
0 \\
1  
\end{array}
\right],
\end{eqnarray*}
where $L$ and $R$ refer to the left and right chirality state, respectively.  

For the general setting, the time evolution of the walk is determined by a $2 \times 2$ unitary matrix, $U$, where
\begin{align*}
U =
\left[
\begin{array}{cc}
a & b \\
c & d
\end{array}
\right],
\end{align*}
with $a, b, c, d \in \mathbb C$ and $\CM$ is the set of complex numbers. The matrix $U$ rotates the chirality before the displacement, which defines the dynamics of the walk. To describe the evolution of our model, we divide $U$ into two matrices:
\begin{eqnarray*}
P =
\left[
\begin{array}{cc}
a & b \\
0 & 0 
\end{array}
\right], 
\quad
Q =
\left[
\begin{array}{cc}
0 & 0 \\
c & d 
\end{array}
\right],
\end{eqnarray*}
with $U = P + Q$. The important point is that $P$ (resp. $Q$) represents that the walker moves to the left (resp. right) at position $x$ at each time step. 

The {\it Hadamard walk} is determined by the Hadamard gate $U = H$:
\begin{eqnarray*}
H=\frac{1}{\sqrt2}
\left[
\begin{array}{cc}
1 & 1 \\
1 &-1 
\end{array}
\right].
\end{eqnarray*}
The walk is intensively investigated in the study of the QW.

In the present paper, we focus on the Hadamard walk and take $\varphi_{\ast} = {}^T [1/\sqrt{2},i/\sqrt{2}]$ as the initial qubit state, where $T$ is the transpose operator. Then the probability distribution of the Hadamard walk starting from $\varphi_{\ast}$ at the origin is symmetric.

Let $\Xi_{n} (l,m)$ denote the sum of all paths starting from the origin in the trajectory consisting of $l$ steps left and $m$ steps right at time $n$ with $n=l+m$. For example, 
\begin{align*}
\Xi_2 (1,1) &= Q P + P Q, \\
\Xi_4 (2,2) &= Q^2 P^2 + P^2 Q^2 + Q P Q P + P Q P Q + P Q^2 P + Q P^2 Q. 
\end{align*}
The probability that our quantum walker is in position $x$ at time $n$ starting from the origin with $\varphi_{\ast} (={}^T [1/\sqrt{2},i/\sqrt{2}])$ is defined by 
\begin{align*}
P (X_{n} =x) = || \Xi_{n}(l, m) \> \varphi_{\ast} ||^2,
\end{align*}
where $n=l+m$ and $x=-l+m$.

Let $\Psi^{x \to y} _{n} (k)$ denote the sum of all paths that the quantum walker, which starts at position $x$ and reaches position $y$ at time $n$, spends exactly $k$ intervals of time to the right of the origin. 

Let $\Psi^x _{n} (k)$ be the sum of all paths that the quantum walker starting from $x$ spends exactly $k$ intervals of time, up to time $n$ to the right of the origin, i.e., $\Psi^x _{n} (k) = \sum_{y=-n+x}^{n+x} \Psi^{x \to y} _{n} (k)$. For example, 
\begin{align*}
\Psi^0 _2 (0)
&= P^2 + Q P, \quad \Psi^0 _2 (1) = O, \quad \Psi^0 _2 (2) = Q^2 + P Q, 
\\
\Psi^1 _2 (0) 
&= O, \quad \Psi^1 _2 (1) = P^2, \quad  \Psi^1 _2 (2) = Q^2 + P Q + Q P,
\end{align*}
where $O$ is the $2 \times 2$ zero matrix.

Let $\Gamma _{n} (k)$ be the sum of all paths that the quantum walker starting from the origin and returning to this point spends exactly $k$ intervals of time, up to time $n$ to the right of the origin, i.e., $\Gamma _{n} (k) = \Psi^{0 \to 0} _{n} (k)$. For example, 
\begin{align*}
\Gamma_2 (0)
&= Q P, \quad  \Gamma_2 (1) = O, \quad \Gamma_2 (2) = P Q, 
\\
\Gamma_4 (0)
&= \Gamma_4 (1) = \Gamma_4 (3) = \Gamma_4 (4) = O, \quad \Gamma_4 (2) = Q P^2 Q + P Q^2 P.
\end{align*}
The next section will give the generating functions of $\Psi^x _{n} (k)$ and $\Gamma_{n} (k)$. 

\section{Our results}
In this section, we first present the generating function of $\Psi^x _{n} (k)$  for the Hadamard walk. To do so, we introduce $R$ and $S$ as follows:
\begin{align*}
R =
\frac{1}{\sqrt{2}}
\left[
\begin{array}{cc}
1 & -1 \\
0 & 0 
\end{array}
\right], 
\quad
S =
\frac{1}{\sqrt{2}}
\left[
\begin{array}{cc}
0 & 0 \\
1 & 1 
\end{array}
\right].
\end{align*} 
We should note that
\begin{align*}
P =
\frac{1}{\sqrt{2}}
\left[
\begin{array}{cc}
1 & 1 \\
0 & 0 
\end{array}
\right], 
\quad
Q =
\frac{1}{\sqrt{2}}
\left[
\begin{array}{cc}
0 & 0 \\
1 & -1 
\end{array}
\right].
\end{align*} 
Then $P, Q, R$ and $S$ form an orthonormal basis of the vector space of complex $2 \times 2$ matrices with respect to the trace inner product $\langle A | B \rangle = $ tr$(A^{\ast}B)$, where $\ast$ means the adjoint operator. Thus we have the following expression: 
\begin{align*} 
\Psi^x _n (k) = p^x _n (k) P + q^x _n (k) Q + r^x _n (k) R + s^x _n (k) S. 
\end{align*} 
For $u \in \{p,q,r,s\}$, we introduce the generating function 
\begin{align*}
\tilde{u}^{x} (z, t) 
= \sum_{k=0}^{\infty} \sum_{n=1}^{\infty} u^x _n (k) z^n t^k. 
\end{align*}
Moreover, for $u \in \{p,q,r,s\}$, we let
\begin{align*}
\bar{u}^{x} (z, t) 
= \tilde{u}^{x} (z, t) +  \tilde{u}^{x} (-z, t) + \tilde{u}^{x} (z, -t) +  \tilde{u}^{x} (-z, -t).
\end{align*}
Remark that for $u \in \{p,q,r,s\}$, we see 
\begin{align*}
\bar{u}^{x} (z, t) = \sum_{k=0}^{\infty} \sum_{n=1}^{\infty} u^x _{2n} (2k) z^{2n} t^{2k}.
\end{align*}
Then we obtain
\begin{thm}
\label{thm1a}
\begin{align*}
\bar{p}^{0} (z, t) 
&= 
\frac{z^2 t^2 \left\{ 2(1-t^2) z^2 + (1 - z^2 t^2) A +(1-z^2) B \right\}}{\sqrt{2}(1-z^2)(1-z^2t^2)\left\{ -1 + (1+A) t^2 + B \right\}},
\\
\bar{r}^{0} (z, t) 
&= 
\frac{- z^4 t^2 + ( 1 + A ) (1 - B ) + z^2 \left\{ 1 + (1+A) t^2 - B \right\}}{2 \sqrt{2}(1-z^2)(1-z^2t^2)},
\\
\bar{q}^{0} (z, t) 
&= 
\frac{z^2 t^2 \left[ -z^4 + z^6 t^2 + (1 + A ) \left\{ -1 + (2-t^2)z^2 - (1 - z^2) B \right\} \right]}{\sqrt{2}(1-z^2)(1-z^2t^2)(1 + A ) \left\{1 - (1 - A) t^2 + B \right\}},
\\
\bar{s}^{0} (z, t) 
&= 
\frac{z^4 t^2 \left\{ (1 - z^4 t^4) A + (1-z^2)(1+z^2t^2) B + (1 - z^2 t^2) A B + (1-z^2) B^2 \right\}}{\sqrt{2}(1-z^2)(1-z^2t^2)(1 + A)(-1+z^2t^2 + B) \left\{ 1 - (1 - A)t^2 + B) \right\}}, 
\end{align*}
where $A=\sqrt{1+z^4}$ and $B = \sqrt{1+z^4 t^4}.$ 
\end{thm}
The proof is given in Section 4. From this theorem, 
\begin{align*}
\bar{p}^{0} (z, t) 
&= \frac{1}{\sqrt{2}} z^2 + \frac{3 - t^2}{2 \sqrt{2}} z^4 + \frac{6 - t^2 - t^4}{4 \sqrt{2}} z^6 + \frac{11 - 2 t^2 + t^4 - 2 t^6}{8 \sqrt{2}} z^8 + \cdots,   
\\
\bar{r}^{0} (z, t) 
&= \frac{t^2}{\sqrt{2}} z^2 + \frac{t^2+t^4}{2 \sqrt{2}} z^4 + \frac{3 t^2 - t^4 + 2 t^6}{4 \sqrt{2}} z^6 + \frac{6 t^2 - t^4 - 2 t^6 + 5 t^8}{8 \sqrt{2}} z^8 + \cdots,   
\\
\bar{q}^{0} (z, t) 
&= - \frac{t^2}{\sqrt{2}} z^2 + \frac{t^2 - 3 t^4}{2 \sqrt{2}} z^4 + \frac{t^2 + t^4 - 6 t^6}{4 \sqrt{2}} z^6 + \frac{2 t^2 - t^4 + 2 t^6 - 11 t^8}{8 \sqrt{2}} z^8 + \cdots,
\\
\bar{s}^{0} (z, t) 
&= \frac{1}{\sqrt{2}} z^2 + \frac{1 + t^2}{2 \sqrt{2}} z^4 + \frac{2 - t^2 + 3 t^4}{4 \sqrt{2}} z^6 + \frac{5 - 2 t^2 - t^4 + 6 t^6}{8 \sqrt{2}} z^8 + \cdots.
\end{align*}
For example, we can confirm that the above expression holds for $2n=4$ by using a direct computation as follows: 
\begin{align*}
\Psi^0 _4 (0) 
&= P^4 + Q P^3 + P Q P^2 + Q^2 P^2 + P^2 Q P + QPQP 
= \frac{1}{2\sqrt{2}} (3 P + S),
\\
\Psi^0 _4 (2) 
&= Q P^2 Q + P Q^2 P + P^3 Q + Q^3 P 
= \frac{1}{2\sqrt{2}} (- P +  R  + Q + S),
\\
\Psi^0 _4 (4) 
&= Q^4 + P Q^3 + Q P Q^2 + P^2 Q^2 + Q^2 P Q + PQPQ 
= \frac{1}{2\sqrt{2}} (R - 3 Q).
\end{align*}
Moreover we see that
\begin{align*}
p^0 _{4} (0) &= \frac{3}{2 \sqrt{2}}, \quad
p^0 _{4} (2) = - \frac{1}{2 \sqrt{2}}, \quad
p^0 _{4} (4) = 0, \\
r^0 _{4} (0) &= 0, \quad
r^0 _{4} (2) = \frac{1}{2 \sqrt{2}}, \quad
r^0 _{4} (4) = \frac{1}{2 \sqrt{2}}, \\
q^0 _{4} (0) &= 0, \quad
q^0 _{4} (2) = \frac{1}{2 \sqrt{2}}, \quad
q^0 _{4} (4) = - \frac{3}{2 \sqrt{2}}, \\
s^0 _{4} (0) &= \frac{1}{2 \sqrt{2}}, \quad
s^0 _{4} (2) =  \frac{1}{2 \sqrt{2}}, \quad
s^0 _{4} (4) = 0.
\end{align*}
The measure that the quantum walker starting from $x$ spends exactly $k$ intervals of time, up to time $n$ to the right of the origin is defined by 
\begin{align*}
Q \left( A^x _{n} = k \right) = || \Psi^x _{n} (k) \> \varphi_{\ast} ||^2,
\end{align*}
where $A^x _{n}$ is the sojourn time. Noting the initial qubit state $\varphi_{\ast} ={}^T [1/\sqrt{2},i/\sqrt{2}]$, we have
\begin{align}
Q \left( A^0 _{2n} = 2k \right) = \frac{1}{2} \left\{ p^0 _{2n} (2k)^2 +  r^0 _{2n} (2k)^2 + q^0 _{2n} (2k)^2 + s^0 _{2n} (2k)^2 \right\} \quad (k=0,1, \ldots, n).
\label{satomiJuly1}
\end{align}
By using this, in $2n=4$ case we get
\begin{align*}
Q \left( A^0 _{4} = 0 \right) = \frac{5}{8}, \quad Q \left( A^0 _{4} = 2 \right) = \frac{2}{8}, \quad Q \left( A^0 _{4} = 4 \right) = \frac{5}{8}.
\end{align*}
So after the normalization, we have the corresponding probability measure as follows:
\begin{align*}
P \left( A^0 _{4} = 0 \right) = \frac{5}{12}, \quad P \left( A^0 _{4} = 2 \right) = \frac{2}{12}, \quad P \left( A^0 _{4} = 4 \right) = \frac{5}{12}.
\end{align*}
\par
We should remark that (\ref{satomiJuly1}) can be extended for general initial qubit state $\varphi_{\ast} ={}^T [\alpha,\beta]$ with $|\alpha|^2 + |\beta|^2 =1$ as follows: 
\begin{align}
Q \left( A^0 _{2n} = 2k \right) 
&= 
\frac{1}{2} \left\{ p^0 _{2n} (2k)^2 +  r^0 _{2n} (2k)^2 + q^0 _{2n} (2k)^2 + s^0 _{2n} (2k)^2 \right\}  
\nonumber 
\\
&+ \left\{ p^0 _{2n} (2k) r^0 _{2n} (2k) +  q^0 _{2n} (2k) s^0 _{2n} (2k) \right\} \left( |\alpha|^2 - |\beta|^2 \right) 
\nonumber
\\
&+ \frac{1}{2} \left\{ p^0 _{2n} (2k)^2 - r^0 _{2n} (2k)^2 - q^0 _{2n} (2k)^2 + s^0 _{2n} (2k)^2 \right\} \left( \alpha \overline{\beta} +  \overline{\alpha} \beta \right).
\label{satomiJuly2}
\end{align}
On the other hand, it is known that a necessary and sufficient condition of initial qubit state for the symmetry of distribution of the Hadamard walk is $|\alpha|=|\beta|=1/\sqrt{2}$ and $\alpha \overline{\beta} +  \overline{\alpha} \beta = 0$ (see Theorem 6 in \cite{Konno2005}). Thus, for this symmetric Hadamard walk, we see that (\ref{satomiJuly2}) becomes (\ref{satomiJuly1}).

Let $c^{(n)} (k)$ denote the probability that in the time interval from 0 to $n$ the quantum walker starting from the origin spends $k$ time units on the positive side and $n-k$ time units on the negative side and $m^{(n)}$ be the probability measure, i.e., 
\begin{align*}
m^{(n)} = \sum_{k=0}^{n} c^{(n)} (k) \> \delta_{k},
\end{align*}
where $\delta_{k}$ is the Dirac measure at $k$. Therefore Theorem {\rmfamily \ref{thm1a}} yields 
\begin{align*}
m^{(2)}
&= \frac{1}{2} \delta_{0} + \frac{1}{2} \delta_{2}, 
\quad
m^{(4)}
= \frac{5}{12} \delta_{0} + \frac{2}{12} \delta_{2} + \frac{5}{12} \delta_{4}, 
\quad
m^{(6)}
= \frac{10}{26} \delta_{0} + \frac{3}{26} \delta_{2} + \frac{3}{26} \delta_{4} + \frac{10}{26} \delta_{6}, 
\\
m^{(8)}
&= \frac{73}{196} \delta_{0} + \frac{24}{196} \delta_{2} + \frac{2}{196} \delta_{4} + \frac{24}{196} \delta_{6} + \frac{73}{196} \delta_{8}.
\end{align*}
On the other hand, in the classical counterpart, it is well known (see Chapter III in Feller \cite{Feller1968}) that the measure is
\begin{align*}
m^{(2n:c)} = \sum_{k=0} ^{n} c^{(2n)} (2k) \> \delta_{2k} \quad (k=0,1, \ldots, n),
\end{align*}
where
\begin{align*}
c^{(2n)} (2k) = \left( \frac{1}{2} \right)^{2n} {2k \choose k} {2(n-k) \choose n-k} \quad (k=0,1, \ldots, n).
\end{align*}
It is called the discrete arc sine distribution. For example, 
\begin{align*}
m^{(2:c)}
&= \frac{1}{2} \delta_{0} + \frac{1}{2} \delta_{2}, 
\quad
m^{(4:c)}
= \frac{3}{8} \delta_{0} + \frac{2}{8} \delta_{2} + \frac{3}{8} \delta_{4}, 
\quad
m^{(6:c)}
= \frac{5}{16} \delta_{0} + \frac{3}{16} \delta_{2} + \frac{3}{16} \delta_{4} + \frac{5}{26} \delta_{6}, 
\\
m^{(8:c)}
&= \frac{35}{128} \delta_{0} + \frac{20}{128} \delta_{2} + \frac{18}{128} \delta_{4} + \frac{20}{128} \delta_{6} + \frac{35}{128} \delta_{8}.
\end{align*}
The central term is always smallest for both cases, however we find that the corresponding term of the QW version is smaller than that of the classical random walk for small $n$ as follows: 
\begin{align*}
\frac{2}{15} < \frac{2}{8} \quad (2n=4), \quad \frac{3}{26} < \frac{3}{16} \quad (2n=6), \quad \frac{2}{196} < \frac{18}{128} \quad (2n=8). 
\end{align*}
Moreover we should remark that the generating function of the classical random walk $S^{0} _{2n}$ starting from the origin at time $2n$ is given by
\begin{align*}
\sum_{k=0}^{\infty} \sum_{n=0}^{\infty} P(S^{0} _{2n} = 2k) z^{2n} t^{2k} 
=  \frac{1}{\sqrt{1-z^2} \sqrt{1-z^2t^2}}.
\end{align*}
See Fujita and Yor \cite{FujitaYor2007} for its related results.

As in the previous case, we next consider the following generating function of $\Gamma_{2n} (2k)$ for the Hadamard walk:  
\begin{align*}
\bar{\Gamma} (z, t) = \sum_{k=0}^{\infty} \sum_{n=1}^{\infty} \Gamma_{2n} (2k)  z^{2n} t^{2k}.
\end{align*}
Then we have
\begin{thm}
\label{thm2a}
\begin{align*}
\bar{\Gamma} (z, t) = \frac{1}{C}
\left[
\begin{array}{cc}
- (z^2 - A)(1+z^2t^2-B) & 1+z^2t^2-B \\
-1 - z^2 + A & - (-1 - z^2 + A)(z^2t^2-B)
\end{array}
\right],
\end{align*}
where $A=\sqrt{1+z^4}, \> B = \sqrt{1+z^4 t^4}$ and $C = - 1 - (z^2 - A)(z^2t^2-B).$ 
\end{thm}
The proof will appear in Section 5. By this theorem, we see 
\begin{align*}
\bar{\Gamma} (z, t) 
&= \frac{1}{2} 
\left[
\begin{array}{cc}
t^2 & - t^2\\
1 & 1 
\end{array}
\right] z^2
+
\frac{1}{4} 
\left[
\begin{array}{cc}
- t^2 & - t^2 \\
t^2 & - t^2
\end{array}
\right] z^4
+
\frac{1}{8} 
\left[
\begin{array}{cc}
- (t^4 + t^6) & t^4 + t^6 \\
-(1 + t^2) & -(1 + t^2)
\end{array}
\right] z^6
\\
&+
\frac{1}{16} 
\left[
\begin{array}{cc}
t^2 + t^4 + t^6 & t^2 + t^4 + t^6 \\
-(t^2 + t^4 + t^6) & t^2 + t^4 + t^6
\end{array}
\right] z^8
\\
&+
\frac{1}{32} 
\left[
\begin{array}{cc}
t^2 + t^6 + 2 t^8 + 2 t^{10} & - (t^2 + t^6 + 2 t^8 + 2 t^{10}) \\
2 + 2 t^2 + t^4 + t^6 & 2 + 2 t^2 + t^4 + t^6
\end{array}
\right] z^{10}
+ \cdots.
\end{align*}
For example, we can confirm that the above expression holds for $2n =2, \> 4$ by using a direct computation in the following way: 
\begin{align*}
\Gamma_2 (0) 
&= QP =
\frac{1}{2} 
\left[
\begin{array}{cc}
0 & 0 \\
1 & 1
\end{array}
\right], 
\quad
\Gamma_2 (2) 
= P Q =
\frac{1}{2} 
\left[
\begin{array}{cc}
1 & - 1 \\
0 & 0
\end{array}
\right], 
\\
\Gamma_4 (0)
&= \Gamma_4 (4) = O, \quad 
\Gamma_4 (2) = Q P^2 Q + P Q^2 P =
\frac{1}{4} 
\left[
\begin{array}{cc}
- 1 & - 1 \\
1 & - 1
\end{array}
\right].
\end{align*}
The measure that the quantum walker starting from the origin and returning to this point spends exactly $k$ intervals of time, up to time $n$ to the right of the origin is defined by
\begin{align*}
Q \left( B _{n} = k \right) = || \Gamma_{n} (k) \> \varphi_{\ast} ||^2, 
\end{align*}
where $B _{n}$ denotes the sojourn time. Let $\Gamma_{2n}^{(i,j)} (2k)$ denote the $(i,j)$ component of $\Gamma_{2n} (2k)$ for $i,j \in \{1,2\}.$ Noting the initial qubit state $\varphi_{\ast} ={}^T [1/\sqrt{2},i/\sqrt{2}]$, we have
\begin{align}
Q (B_{2n} = 2k) = \frac{1}{2} \sum_{i=1}^2 \sum_{j=1}^2 \Gamma_{2n}^{(i,j)} (2k)^2.
\label{satomiJuly3}
\end{align}
Thus
\begin{align*}
Q (B_2 = 0) 
&= Q (B_2 = 2) = \frac{1}{4}, \quad Q (B_4 = 0) = Q (B_4 = 4) =0, \>\> Q (B_4 = 2) = \frac{1}{8}, 
\\
Q (B_6 = 0) 
&= Q (B_6 = 2) = Q (B_6 = 4) = Q (B_6 = 6) = \frac{1}{64}, \ldots.
\end{align*}
After the normalization, the corresponding probability measure becomes
\begin{align*}
P (B_2 = 0) 
&= P (B_2 = 2) = \frac{1}{2}, \quad P (B_4 = 0) = P (B_4 = 4) =0, \>\> P (B_4 = 2) = 1, 
\\
P (B_6 = 0) 
&= P (B_6 = 2) = P (B_6 = 4) = P (B_6 = 6) = \frac{1}{4}, \ldots.
\end{align*}
\par
We note that (\ref{satomiJuly3}) can be extended for general initial qubit state $\varphi_{\ast} ={}^T [\alpha,\beta]$ with $|\alpha|^2 + |\beta|^2 =1$ as follows: 
\begin{align}
Q (B_{2n} = 2k) 
&= 
\left\{ \Gamma_{2n}^{(1,1)} (2k)^2 + \Gamma_{2n}^{(2,1)} (2k)^2 \right\} |\alpha|^2 + \left\{ \Gamma_{2n}^{(1,2)} (2k)^2 + \Gamma_{2n}^{(2,2)} (2k)^2 \right\} |\beta|^2 
\nonumber 
\\
&+ \left\{ \Gamma_{2n}^{(1,1)} (2k) \Gamma_{2n}^{(1,2)} (2k) + \Gamma_{2n}^{(2,1)} (2k) \Gamma_{2n}^{(2,2)} (2k) \right\} \left( \alpha \overline{\beta} +  \overline{\alpha} \beta \right).
\label{satomiJuly4}
\end{align}
On the other hand, a necessary and sufficient condition for initial qubit state of symmetric Hadamard walk is $|\alpha|=|\beta|=1/\sqrt{2}$ and $\alpha \overline{\beta} +  \overline{\alpha} \beta = 0$ (see Theorem 6 in \cite{Konno2005}). So, in the symmetric walk, we observe that (\ref{satomiJuly4}) is reduced to (\ref{satomiJuly3}).

Let $\mu^{(n)}$ denote the probability measure at time $n$. Therefore we have 
\begin{align*}
\mu^{(2)}
&= \frac{1}{2} \delta_{0} + \frac{1}{2} \delta_{2}, 
\quad
\mu^{(4)}
= \delta_{2}, 
\quad
\mu^{(6)}
= \frac{1}{4} \delta_{0} + \frac{1}{4} \delta_{2} + \frac{1}{4} \delta_{4} + \frac{1}{4} \delta_{6}, 
\\
\mu^{(8)}
&= \frac{1}{3} \delta_{2} + \frac{1}{3} \delta_{4} + \frac{1}{3} \delta_{6}, 
\quad
\mu^{(10)}
= \frac{2}{10} \delta_{0} + \frac{2}{10} \delta_{2} + \frac{1}{10} \delta_{4} + \frac{1}{10} \delta_{6} + \frac{2}{10} \delta_{8} + \frac{2}{10} \delta_{10}, 
\\
\quad
\mu^{(12)}
&= \frac{1}{5} \delta_{2} + \frac{1}{5} \delta_{4} + \frac{1}{5} \delta_{6} + \frac{1}{5} \delta_{8} + \frac{1}{5} \delta_{10}, 
\\
\mu^{(14)}
&= \frac{25}{152} \delta_{0} + \frac{25}{152} \delta_{2} + \frac{13}{152} \delta_{4} + \frac{13}{152} \delta_{6} + \frac{13}{152} \delta_{8} + \frac{13}{152} \delta_{10} + \frac{25}{152} \delta_{12} + \frac{25}{152} \delta_{14}.
\end{align*}
In particular, we consider $4n$ case. Then we see
\begin{align*}
\sum_{k=0}^{\infty} \sum_{n=1}^{\infty} \Gamma_{4n} (2k)  z^{4n} t^{2k}
&= \frac{1}{2} \left\{ \bar{\Gamma} (z, t) + \bar{\Gamma} (iz, t) \right\}
\\
&= \frac{1 - (1 - A) t^2 - B}{2(1-t^2)} 
\left[
\begin{array}{cc}
- 1 & - 1 \\
1 & - 1
\end{array}
\right]
\\
&=
\frac{1}{2} \sum_{n=1}^{\infty} b_n (t^2 + t^4 + \cdots + t^{4n-2}) 
\left[
\begin{array}{cc}
- 1 & - 1 \\
1 & - 1
\end{array}
\right]
z^{4n},
\end{align*}
where 
\begin{align*}
\sqrt{1+z} = \sum_{n=0}^{\infty} b_n z^n, 
\end{align*}
with $b_0 = 1, \> b_1 = 1/2, \> b_2 = -1/8, \ldots.$ Here the second equality comes from Theorem {\rmfamily \ref{thm2a}}. Thus we have
\begin{cor}
\begin{align*}
\mu^{(4n)} = \frac{1}{2n-1} \sum_{k=1}^{2n-1} \delta_{2k} \quad (n=1,2, \ldots). 
\end{align*}
\end{cor}
That is, for the $4n$ case, the measure becomes the uniform distribution on the region $\{2, 4, \ldots, 4n-2 \}$. On the other hand, in the classical counterpart, it is known that the measure is the uniform distribution for any $n \ge 1$, i.e., 
\begin{align*}
\mu^{(2n:c)} = \frac{1}{n+1} \sum_{k=0}^n \delta_{2k}. 
\end{align*}
The result is called the equidistribution theorem (see Chapter III in Feller \cite{Feller1968}).

\section{Proof of Theorem {\rmfamily \ref{thm1a}}}
In this section, we prove Theorem {\rmfamily \ref{thm1a}} by using a path counting approach. Let $\tilde{\Psi}^x (k;z) = \sum_{n=1}^{\infty} \Psi^x _{n} (k) z^n.$ For example, we consider $k = 2$ case. For $x \ge 3$, the definition gives   
\begin{align*}
\tilde{\Psi}^x (2;z) = \Psi^x _{2} (2) z^2 = U z^2.
\end{align*}
For $x=1, 2$, 
\begin{align*}
\Psi^x _{n+1} (2) = \Psi^{x-1} _{n} (1) P + \Psi^{x+1} _{n} (1) Q. 
\end{align*}
Thus 
\begin{align*}
\tilde{\Psi}^x (2;z) = z \tilde{\Psi}^{x-1} (1;z) P + z \tilde{\Psi}^{x+1} (1;z) Q.
\end{align*}
For $x=0$, 
\begin{align*}
\tilde{\Psi}^0 (2;z) = z \tilde{\Psi}^{-1} (2;z) P + z \tilde{\Psi}^{1} (1;z) Q.
\end{align*}
For $x \le -1$,
\begin{align*}
\tilde{\Psi}^x (2;z) = z \tilde{\Psi}^{x-1} (2;z) P + z \tilde{\Psi}^{x+1} (2;z) Q.
\end{align*}
For other $k$ cases, we can compute $\tilde{\Psi}^x (k;z)$ in a similar fashion. Put $a_{x,k} = \tilde{\Psi}^x (k;z)$. So we have the following table: 
\par
\
\par
\begin{center}
\begin{tabular}{c|cccc}
$x \backslash k$  & 0 & 1 & 2 & 3  \\ \hline
5 & $O$ & $z U$ & $z^2 U^2$ & $z^3 U^3$  \\
4 & $O$ & $z U$ & $z^2 U^2$ & $z^3 U^3$ \\
3 & $O$ & $z U$ & $z^2 U^2$ & $z a_{2,2}P + z a_{4,2}Q$ \\
2 & $O$ & $z U$ & $z a_{1,1}P + z a_{3,1}Q$ & $z a_{1,2}P + z a_{3,2}Q$ \\
1 & $O$ & $z a_{0,0}P + z U$ & $z a_{0,1}P + z a_{2,1}Q$ & $z a_{0,2}P + z a_{2,2}Q$ \\
0 & $z a_{-1,0}P + z P$ & $z a_{-1,1}P + z Q$ & $z a_{-1,2}P + z a_{1,1}Q$ & $z a_{-1,3}P + z a_{1,2}Q$ \\
-1 & $z a_{-2,0}P + z a_{0,0} Q + z U$ & $z a_{-2,1}P + z a_{0,1}Q$ & $z a_{-2,2}P + z a_{0,1}Q$ & $z a_{-2,3}P + z a_{0,3}Q$ \\
-2 & $z a_{-3,0}P + z a_{-1,0} Q + z U$ & $z a_{-3,1}P + z a_{-1,1}Q$ & $z a_{-3,2}P + z a_{-1,1}Q$ & $z a_{-3,3}P + z a_{-1,3}Q$ \\
-3 & $z a_{-4,0}P + z a_{-2,0} Q + z U$ & $z a_{-4,1}P + z a_{-2,1}Q$ & $z a_{-4,2}P + z a_{-2,1}Q$ & $z a_{-4,3}P + z a_{-4,3}Q$ \\
-4 & $z a_{-5,0}P + z a_{-3,0} Q + z U$ & $z a_{-5,1}P + z a_{-3,1}Q$ & $z a_{-5,2}P + z a_{-3,1}Q$ & $z a_{-5,3}P + z a_{-3,3}Q$ 
\label{pqrs}
\end{tabular}
\end{center}
For example, $a_{0,0} = z a_{-1,0}P + z P, \> a_{-1,0} = z a_{-2,0}P + z a_{0,0} Q + z U$. We should remark that $\Psi^x _n (k) = p^x _n (k) P + q^x _n (k) Q + r^x _n (k) R + s^x _n (k) S$. Then for $u \in \{p,q,r,s\}$, we put
\begin{align*}
\tilde{u}^{x} (k; z) = \sum_{n=1}^{\infty} u^x _n (k) z^n. 
\end{align*}
First we consider $x=0$ case. From the above table, we see $a_{0,0} = z a_{-1,0}P + z P$. So
\begin{align*}
\tilde{\Psi}^0 (0;z)
&= \tilde{p}^0 (0; z) P + \tilde{q}^0 (0; z) Q + \tilde{r}^0 (0; z) R + \tilde{s}^0 (0; z) S
\\
&= z \left\{ \tilde{p}^{-1} (0; z) P + \tilde{q}^{-1} (0; z) Q + \tilde{r}^{-1} (0; z) R + \tilde{s}^{-1} (0; z) S \right\} + z P
\\
&= z \left\{ a \tilde{p}^{-1} (0; z) + c \tilde{r}^{-1} (0; z) + 1 \right\} P + z \left\{ c \tilde{q}^{-1} (0; z) + a \tilde{s}^{-1} (0; z) \right\} S.
\end{align*}
The third equality comes from the following algebra:
\begin{center}
\begin{tabular}{c|cccc}
  & $P$ & $Q$ & $R$ & $S$  \\ \hline
$P$ & $aP$ & $bR$ & $aR$ & $bP$  \\
$Q$ & $cS$ & $dQ$& $cQ$ & $dS$ \\
$R$ & $cP$ & $dR$& $cR$ & $dP$ \\
$S$ & $aS$ & $bQ$ & $aQ$ & $bS$ 
\label{pqrs}
\end{tabular}
\end{center}
where $PQ=bR$, for example. Thus 
\begin{align*}
\tilde{p}^0 (0; z) 
&= z \left\{ a \tilde{p}^{-1} (0; z) + c \tilde{r}^{-1} (0; z) + 1 \right\}, \quad \tilde{q}^0 (0; z) = \tilde{r}^0 (0; z) = 0, 
\\
\tilde{s}^0 (0; z) 
&= z \left\{ c \tilde{q}^{-1} (0; z) + a \tilde{s}^{-1} (0; z) \right\}.
\end{align*}
In a similar way, we obtain
\begin{lem}
\label{lem40}
(i)  
\begin{align*}
\tilde{p}^0 (k; z) 
= 
\left\{
\begin{array}{cl}
\displaystyle{z \left\{ a \tilde{p}^{-1} (k; z) + c \tilde{r}^{-1} (k; z) \right\}} & \mbox{if $k=1,2, \ldots$}, \\
\displaystyle{z \left\{ a \tilde{p}^{-1} (0; z) + c \tilde{r}^{-1} (0; z) + 1 \right\}} & \mbox{if $k=0$},
\end{array}
\right.
\end{align*}
(ii) 
\begin{align*}
\tilde{s}^0 (k; z) 
= z \left\{ c \tilde{q}^{-1} (k; z) + a \tilde{s}^{-1} (k; z) \right\} \quad \mbox{if $k=0,1,2, \ldots$},
\end{align*}
(iii)  
\begin{align*}
\tilde{r}^0 (k; z) 
= 
\left\{
\begin{array}{cl}
\displaystyle{z \left\{ b \tilde{p}^{1} (k-1; z) + d \tilde{r}^{1} (k-1; z) \right\}} & \mbox{if $k=2,3, \ldots$}, \\
\displaystyle{0} & \mbox{if $k=0,1$},
\end{array}
\right.
\end{align*}
(iv)  
\begin{align*}
\tilde{q}^0 (k; z) 
= 
\left\{
\begin{array}{cl}
\displaystyle{z \left\{ d \tilde{q}^{1} (k-1; z) + b \tilde{s}^{1} (k-1; z) \right\}} & \mbox{if $k=2,3, \ldots$}, \\
\displaystyle{z} & \mbox{if $k=1$}, \\
\displaystyle{0} & \mbox{if $k=0$}.
\end{array}
\right.
\end{align*}
\end{lem}
For $u \in \{p,q,r,s\}$, we put
\begin{align*}
\tilde{u}^{x} (z, t) 
= \sum_{k=0}^{\infty} \tilde{u}^x (k; z) t^k
\left( = \sum_{k=0}^{\infty} \sum_{n=1}^{\infty} u^x _n (k) z^n t^k \right). 
\end{align*}
So 
\begin{align*}
\tilde{p}^{0} (z, t) 
&= \sum_{k=0}^{\infty} \tilde{p}^0 (k; z) t^k
= z \sum_{k=1}^{\infty} \left\{ a \tilde{p}^{-1} (k; z) + c \tilde{r}^{-1} (k; z) \right\} t^k + z
\\
&= z \left\{a \tilde{p}^{-1} (z,t) + c \tilde{r}^{-1} (z,t) + 1 \right\}.
\end{align*}
Then we can compute $\tilde{q}^{0} (z, t), \tilde{r}^{0} (z, t)$ and $\tilde{s}^{0} (z, t)$ also. Furthermore we get the corresponding equations for $|x| \ge 1$ cases in a similar way, so we omit the details. Therefore we obtain
\begin{lem}
\label{lem41}
(i) For $x \le -1$, 
\begin{align*}
\tilde{p}^x (z,t) 
&= z \left\{a \tilde{p}^{x-1} (z,t) + c \tilde{r}^{x-1} (z,t) + 1 \right\}, \\
\tilde{r}^x (z,t) 
&= z \left\{b \tilde{p}^{x+1} (z,t) + d \tilde{r}^{x+1} (z,t) \right\}, \\
\tilde{q}^x (z,t) 
&= z \left\{d \tilde{q}^{x+1} (z,t) + b \tilde{s}^{x+1} (z,t) + 1 \right\}, \\
\tilde{s}^x (z,t) 
&= z \left\{c \tilde{q}^{x-1} (z,t) + a \tilde{s}^{x-1} (z,t) \right\}. 
\end{align*}
(ii) 
\begin{align*}
\tilde{p}^0 (z,t) 
&= z \left\{a \tilde{p}^{-1} (z,t) + c \tilde{r}^{-1} (z,t) + 1 \right\}, \\
\tilde{r}^0 (z,t) 
&= z t \left\{b \tilde{p}^{1} (z,t) + d \tilde{r}^{1} (z,t) \right\}, \\
\tilde{q}^0 (z,t) 
&= z t \left\{d \tilde{q}^{1} (z,t) + b \tilde{s}^{1} (z,t) + 1 \right\}, \\
\tilde{s}^0 (z,t) 
&= z \left\{c \tilde{q}^{-1} (z,t) + a \tilde{s}^{-1} (z,t) \right\}. 
\end{align*}
(i) For $x \ge 1$, 
\begin{align*}
\tilde{p}^x (z,t) 
&= z t \left\{a \tilde{p}^{x-1} (z,t) + c \tilde{r}^{x-1} (z,t) + 1 \right\}, \\
\tilde{r}^x (z,t) 
&= z t \left\{b \tilde{p}^{x+1} (z,t) + d \tilde{r}^{x+1} (z,t) \right\}, \\
\tilde{q}^x (z,t) 
&= z t \left\{d \tilde{q}^{x+1} (z,t) + b \tilde{s}^{x+1} (z,t) + 1 \right\}, \\
\tilde{s}^x (z,t) 
&= z t \left\{c \tilde{q}^{x-1} (z,t) + a \tilde{s}^{x-1} (z,t) \right\}. 
\end{align*}
\end{lem}
Let $\triangle = ad -bc$. From this lemma, we immediately get
\begin{cor}
\label{cor42}
(i) For $x \le -2$, 
\begin{align*}
& d \tilde{p}^{x+2} (z,t) 
- \left( \triangle z + \frac{1}{z} \right) \tilde{p}^{x+1} (z,t) + a \tilde{p}^{x} (z,t) - dz + 1 = 0, \\
& d \tilde{r}^{x+2} (z,t) 
- \left( \triangle z + \frac{1}{z} \right) \tilde{r}^{x+1} (z,t) + a \tilde{r}^{x} (z,t) + bz = 0, \\
& d \tilde{q}^{x+2} (z,t) 
- \left( \triangle z + \frac{1}{z} \right) \tilde{q}^{x+1} (z,t) + a \tilde{q}^{x} (z,t) - az + 1 = 0, \\
& d \tilde{s}^{x+2} (z,t) 
- \left( \triangle z + \frac{1}{z} \right) \tilde{s}^{x+1} (z,t) + a \tilde{s}^{x} (z,t) + cz = 0. 
\end{align*}
(i) For $x \ge 0$, 
\begin{align*}
& d \tilde{p}^{x+2} (z,t) 
- \left( \triangle zt + \frac{1}{zt} \right) \tilde{p}^{x+1} (z,t) + a \tilde{p}^{x} (z,t) - dzt + 1 = 0, \\
& d \tilde{r}^{x+2} (z,t) 
- \left( \triangle zt + \frac{1}{zt} \right) \tilde{r}^{x+1} (z,t) + a \tilde{r}^{x} (z,t) + bzt = 0, \\
& d \tilde{q}^{x+2} (z,t) 
- \left( \triangle zt + \frac{1}{zt} \right) \tilde{q}^{x+1} (z,t) + a \tilde{q}^{x} (z,t) - azt + 1 = 0, \\
& d \tilde{s}^{x+2} (z,t) 
- \left( \triangle zt + \frac{1}{zt} \right) \tilde{s}^{x+1} (z,t) + a \tilde{s}^{x} (z,t) + czt = 0. 
\end{align*}
\end{cor}
Let $\phi (z) = (- \triangle ) z^2 + (a+d) z -1$ and 
\begin{align*}
\lambda_{\pm} (z) = \frac{\triangle z^2 + 1 \mp \sqrt{\triangle^2 z^4 + 2 \triangle (1 - 2 |a|^2) z^2 + 1}}{2 \triangle \bar{a} z}.
\end{align*}
Moreover we put $\lambda_1 = \lambda_{-} (zt)$ with $|\lambda_1| < 1$ and $\lambda_2 = \lambda_{+} (z)$ with $|\lambda_2| > 1.$ Then we have
\begin{lem}
\label{lem43}
\begin{align*}
\tilde{p}^{x} (z,t)
&= 
\left\{
\begin{array}{cl}
\displaystyle{C_1^{(p)} \lambda_1 ^x + \frac{zt(dzt-1)}{\phi (zt)}} & \mbox{if $x \ge 0$}, \\
\displaystyle{C_2^{(p)} \lambda_2 ^x + \frac{z(dz-1)}{\phi (z)}} & \mbox{if $x \le 0$},
\end{array}
\right.
\\
\tilde{r}^{x} (z,t)
&= 
\left\{
\begin{array}{cl}
\displaystyle{C_1^{(r)} \lambda_1 ^x - \frac{b (zt)^2}{\phi (zt)}} & \mbox{if $x \ge 0$}, \\
\displaystyle{C_2^{(r)} \lambda_2 ^x - \frac{b z^2}{\phi (z)}} & \mbox{if $x \le 0$},
\end{array}
\right.
\\
\tilde{q}^{x} (z,t)
&= 
\left\{
\begin{array}{cl}
\displaystyle{C_1^{(q)} \lambda_1 ^x + \frac{zt(azt-1)}{\phi (zt)}} & \mbox{if $x \ge 0$}, \\
\displaystyle{C_2^{(q)} \lambda_2 ^x + \frac{z(az-1)}{\phi (z)}} & \mbox{if $x \le 0$},
\end{array}
\right.
\\
\tilde{s}^{x} (z,t)
&= 
\left\{
\begin{array}{cl}
\displaystyle{C_1^{(s)} \lambda_1 ^x - \frac{c (zt)^2}{\phi (zt)}} & \mbox{if $x \ge 0$}, \\
\displaystyle{C_2^{(s)} \lambda_2 ^x - \frac{c z^2}{\phi (z)}} & \mbox{if $x \le 0$}.
\end{array}
\right.
\end{align*}
\end{lem}
Combining Lemma {\rmfamily \ref{lem41}} (ii) with Lemma {\rmfamily \ref{lem43}} yields
\begin{align*}
C_2^{(p)} + \frac{z(dz-1)}{\phi (z)}
&= z \left\{ a \left( C_2^{(p)} \lambda_2 ^{-1} + \frac{z(dz-1)}{\phi (z)} \right) + c \left( C_2^{(r)} \lambda_2 ^{-1} - \frac{b z^2}{\phi (z)} \right) + 1 \right\}, \\
C_1^{(r)} - \frac{b (zt)^2}{\phi (zt)} 
&= z t \left\{ b \left( C_1^{(p)} \lambda_1 + \frac{zt(dzt-1)}{\phi (zt)} \right) + d \left( C_1^{(r)} \lambda_1 - \frac{b (zt)^2}{\phi (zt)} \right) \right\}.
\end{align*}
From these, we have
\begin{align}
C_1^{(r)} = \frac{bzt \lambda_1}{1 - bzt \lambda_1} \> C_1^{(p)}, \quad C_2^{(r)} = \frac{\lambda_2 - az}{cz} \> C_2^{(p)}.
\label{rumiko}
\end{align}
By Lemma {\rmfamily \ref{lem43}},
\begin{align*}
C_1^{(p)} + \frac{zt(dzt-1)}{\phi (zt)} 
&= C_2^{(p)} + \frac{z(dz-1)}{\phi (z)} (= \tilde{p}^{0} (t)),  
\\
C_1^{(r)} - \frac{b (zt)^2}{\phi (zt)} 
&= C_2^{(r)} - \frac{b z^2}{\phi (z)} (= \tilde{r}^{0} (t)).
\end{align*}
Combining these with (\ref{rumiko}), we can obtain $C_1^{(p)}, C_2^{(p)}, C_1^{(r)}$ and $C_2^{(r)}.$ Therefore we have $\tilde{p}^{0} (z,t)$ and $\tilde{r}^{0} (z,t)$. In a similar way, we get $\tilde{q}^{0} (z,t)$ and $\tilde{s}^{0} (z,t)$. The explicit expressions of $\tilde{p}^{0} (z,t), \> \tilde{q}^{0} (z,t), \> \tilde{r}^{0} (z,t),$ and $\tilde{s}^{0} (z,t)$ are complicated, so we omit them. By using the relation $\bar{u}^{0} (z, t) = \tilde{u}^{0} (z, t) +  \tilde{u}^{0} (-z, t) + \tilde{u}^{0} (z, -t) +  \tilde{u}^{0} (-z, -t)$ for $u \in \{p,q,r,s\}$, we have the desired conclusion.

\section{Proof of Theorem {\rmfamily \ref{thm2a}}}
In this section, we prove Theorem {\rmfamily \ref{thm2a}} by using a path counting approach as in the proof of Theorem {\rmfamily \ref{thm1a}}. From the definition of $\Gamma_{2n} (2k)$, we easily get 
\begin{lem}
\label{lem5a}
For $k=0,1, \ldots, n$, we have
\begin{align*}
\Gamma_{2n} (2k)
&= I_{\{1,2, \ldots, n \}} (k) \times \sum_{r=1}^k \Gamma_{2n- 2r} (2k-2r) \Gamma_{2r} (2r) 
\\
&+ I_{\{0,1, \ldots, n-1 \}} (k) \times \sum_{r=1}^{n-k} \Gamma_{2n- 2r} (2k) \Gamma_{2r} (0),
\end{align*}
where $I_A (k)$ is the indicator function of a set $A$ and 
\begin{align*}
\Gamma_{2r} (2r) 
= \frac{a_{2r-1}}{2} 
\left[
\begin{array}{cc}
-1 & 1 \\
0 & 0
\end{array}
\right],
\quad 
\Gamma_{2r} (0) 
= \frac{a_{2r-1}}{2} 
\left[
\begin{array}{cc}
0 & 0 \\
-1 & -1
\end{array}
\right] \quad (r \ge 1)
\end{align*}
with
\begin{align}
\sum_{n=1}^{\infty} a_n z^n = \frac{-1-z^2 + \sqrt{1+z^4}}{z}.
\label{mon1}
\end{align}
\end{lem}
The proof is essentially the same as that of Proposition 3.1 in \cite{Konno2009}, so we omit it.
We put 
\begin{align*}
\bar{\Gamma} (z, t) = \sum_{n=1}^{\infty} \sum_{k=0} ^{n} \Gamma_{2n} (2k) t^{2k} z^{2n}. 
\end{align*}
By using Lemma {\rmfamily \ref{lem5a}}, we have
\begin{align}
\bar{\Gamma}_{2n} (t) = \left( \bar{\Gamma}_{2n} (t) + I \right) X, 
\label{mon2}
\end{align}
where $I$ is the $2 \times 2$ identity matrix and 
\begin{align}
X = \sum_{r=1}^{\infty} \left\{ \Gamma_{2r} (2r) (zt)^{2r} + \Gamma_{2r} (0) z^{2r} \right\}.
\label{mon3}
\end{align}
Then (\ref{mon2}) yields 
\begin{align}
\bar{\Gamma}_{2n} (z, t) = X (I -X)^{-1}.
\label{mon4}
\end{align}
From (\ref{mon1}) and (\ref{mon3}), we obtain 
\begin{align}
X = \frac{1}{2}
\left[
\begin{array}{cc}
1 + (zt)^2 - \sqrt{1+(zt)^4} & -1 - (zt)^2 + \sqrt{1+(zt)^4}  \\
1 + z^2 - \sqrt{1+z^4} & 1 + z^2 - \sqrt{1+z^4}
\end{array}
\right].
\label{mon5}
\end{align}
Therefore combining (\ref{mon4}) with (\ref{mon5}) gives the desired conclusion.


\par
\
\par\noindent
{\bf Acknowledgment.} This work was partially supported by the Grant-in-Aid for Scientific Research (C) of Japan Society for the Promotion of Science (Grant No. 21540118).
\par
\
\par

\begin{small}
\bibliographystyle{jplain}

\begin{thebibliography}{99}

\bibitem{Feller1968}
Feller, W.:
An Introduction to Probability Theory and Its Applications, Vol. I, Wiley, New York (1968)


\bibitem{FujitaYor2007}
Fujita, T., Yor, M.:
On the remarkable distributions of maxima of some fragments of the standard reflecting random walk and Brownian motion. 
Probab. Math. Statist. {\bf 27}, 89--104 (2007)


\bibitem{Kempe2003} 
Kempe, J.: 
Quantum random walks - an introductory overview. Contemporary Physics {\bf 44},  307--327 (2003) 


\bibitem{Kendon2007} 
Kendon, V.: 
Decoherence in quantum walks - a review. Math. Struct. in Comp. Sci. {\bf 17}, 1169--1220 (2007)


\bibitem{Konno2005} 
Konno, N.: 
A new type of limit theorems for the one-dimensional quantum random walk.
J. Math. Soc. Japan {\bf 57}, 1179--1195 (2005)


\bibitem{Konno2008b} 
Konno, N.: 
Quantum Walks. In: Quantum Potential Theory, Franz, U., and Sch\"urmann, M., Eds., Lecture Notes in Mathematics: Vol. 1954, pp. 309--452, Springer-Verlag, Heidelberg (2008)


\bibitem{Konno2009} 
Konno, N.: 
One-dimensional discrete-time quantum walks on random environments. 
Quantum Inf. Proc. {\bf 8}, 387--399 (2009)


\bibitem{VAndraca2008} 
Venegas-Andraca, S. E.: 
Quantum Walks for Computer Scientists. Morgan and Claypool (2008)


\end{thebibliography}

\end{small}

\end{document}